\documentclass[letterpaper, 10 pt, conference]{ieeeconf}	

\IEEEoverridecommandlockouts                              

\usepackage{cite}
\usepackage{hyperref} 
\usepackage{graphicx}
\usepackage{psfrag, pstool}
\usepackage{amsmath}
\usepackage{amssymb}
\usepackage{color,soul} 
\newcommand{\figurename}{Fig.}
\usepackage{subfigure}

\title{\LARGE \bf In-silico Feedback Control of a MIMO Synthetic Toggle Switch via Pulse-Width Modulation}

\author{Agostino Guarino$^{1}$, Davide Fiore$^{1}$, Mario di Bernardo$^{1,2}$%
\thanks{$^{1}$Agostino Guarino, Davide Fiore and Mario di Bernardo are with the Department of Electrical Engineering and Information Technology, University of Naples Federico II, Via Claudio 21, 80125 Naples, Italy. E-mail address:
        {\tt\small agostinoguarino@gmail.com, dvd.fiore@gmail.com}}%
\thanks{$^{2}$Mario di Bernardo is also with the Department of Engineering Mathematics, University of Bristol, University Walk, BS8 1TR Bristol, U.K. E-mail address:
        {\tt\small mario.dibernardo@unina.it}}%
}

\begin{document}

\maketitle
\pagestyle{empty}
\thispagestyle{empty}

\begin{abstract}
The synthetic toggle switch, first proposed by Gardner \& Collins \cite{Gardner2000}  is a MIMO control system that can be controlled by varying the concentrations of two inducer molecules, aTc and IPTG, to achieve a desired level of expression of the two genes it comprises.
It has been shown \cite{Lugagne2017} that this can be accomplished through an open-loop external control strategy where the two inputs are selected as mutually exclusive periodic pulse waves of appropriate amplitude and duty-cycle.
In this paper, we use a recently derived average model of the genetic toggle switch subject to these inputs to synthesize new feedback control approaches that adjust the inputs’ duty-cycle in real-time via two different possible strategies, a model-based hybrid PI-PWM approach and a so-called Zero-Average dynamics (ZAD) controller. 
The controllers are validated in-silico via both deterministic and stochastic simulations (SSA) illustrating the advantages and limitations of each strategy.
\end{abstract}

\section{Introduction}
\begin{figure}[!t]
    \centering
	\includegraphics[scale=1]{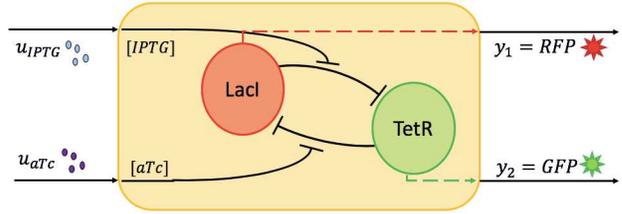}
    \caption{Genetic Toggle Switch as a MIMO System. Inputs are molecules of aTc and IPTG in the growth medium, outputs are fluorescence measures proportional to the concentrations of proteins.}
    \label{fig:GTSMIMO}
\end{figure}

The Genetic Toggle Switch is a genetic network of two mutually repressive genes \cite{Gardner2000} as shown in \figurename~\ref{fig:GTSMIMO}. Each promoter activates the translation of a protein that represses the other promoter. External inducers enhance the proteins production by reducing the repressive effect on the promoter made by the opposite protein.

From a dynamical point of view, the toggle switch is a bistable system, with three equilibria, two stable and one unstable. The two stable equilibria correspond to one gene expression being \emph{high} while the other is \emph{low}, while at the unstable equilibrium neither of the two proteins is fully expressed. Obviously, in-vivo, this situation cannot be maintained for a long time since biological noise drives the circuit onto one of the two stable equilibria.

The problem of controlling the toggle switch dynamics has been the subject of many papers in the literature, and was recently highlighted in \cite{Lugagne2017} as the genetic equivalent of controlling an inverted pendulum. For example, Pulse-Shaping Control \cite{Sootla2010, Sootla2016, Sootla2018} and a Reinforcement Learning control approach \cite{Sootla2013} were both proposed to drive the system from a stable equilibrium to the other. Moreover, Stochastic Motion Planning \cite{Esfahani2013} and Piecewise Linear Switched Control \cite{Chaves2011} were used to stabilize the circuit around its unstable equilibrium. In all these cases though the results are only tested in-silico and no experimental validation is provided.

To the best of our knowledge, the only \emph{in-vivo} control experiment of the toggle switch was reported in \cite{Lugagne2017}. Therein, the authors propose and validate in-vivo different control strategies to stabilize a population of toggle switches in a region surrounding their unstable equilibrium. Firstly, a PI-based solution  is used to close the control loop over a single cell. It is observed that, while the cell being controlled achieves the target region, the others diverge settling down on either of the two stable equilibria.
Secondly, it is shown that, despite closing the loop on a single cell, a bang-bang control approach is  surprisingly able to stabilize the entire population using pulse wave inputs.
Also, inspired by the Kapitza Method for the stabilization of the inverted pendulum \cite{Kapitza1951}, it is reported that an open-loop periodic forcing of the system via two mutually exclusive pulse waves of appropriate frequency and amplitude achieves stabilization of the entire cell population keeping the variance across the population low. 
The problem remains of how to select the required features of the inputs and also of guaranteeing greater robustness given that the strategy is open loop.
In this paper, we leverage our previous modelling work, reported in \cite{Fiore2018}, to synthesize new feedback control strategies able to keep a population of toggle switches at an intermediate level of expression of the genes away from the two stable equilibria. In so doing, we exploit the observation made in \cite{Lugagne2017} that the use of two mutually exclusive pulse wave inputs can enhance coherence in the population. We show that our approach is effective in solving the control problem offering a viable and relatively simple approach to achieve in-vivo external control of a population of genetic toggle switches.

\section{The Toggle Switch as a MIMO Control System}
\label{sec:model}
The deterministic model of the genetic toggle switch that we consider is \cite{Lugagne2017}:
\footnotesize
\begin{align}
\label{eq:transcr_laci}
& \begin{aligned}
\frac{d\, mRNA_{\mathrm{LacI}}}{dt}=\; & \kappa_\mathrm{L}^\mathrm{m0} + \frac{\kappa_\mathrm{L}^\mathrm{m}}{1+ \left( \frac{TetR}{\theta_{\mathrm{TetR}} } \cdot \frac{1}{1 + \left( aTc/\theta_{\mathrm{aTC}} \right)^{\eta_{\mathrm{aTc}}} } \right)^{\eta_{\mathrm{TetR}}} } \\
 & - g_\mathrm{L}^\mathrm{m} \cdot mRNA_{\mathrm{LacI}}
\end{aligned}
\\
\label{eq:trascr_tetr}
& \begin{aligned}
\frac{d\, mRNA_{\mathrm{TetR}}}{dt}=\; & \kappa_\mathrm{T}^\mathrm{m0} + \frac{\kappa_\mathrm{T}^\mathrm{m}}{1+ \left( \frac{LacI}{\theta_{\mathrm{LacI}} } \cdot \frac{1}{1 + \left( IPTG/\theta_{\mathrm{IPTG}} \right)^{\eta_{\mathrm{IPTG}}} } \right)^{\eta_{\mathrm{LacI}}} } \\
 & - g_\mathrm{T}^\mathrm{m} \cdot mRNA_{\mathrm{TetR}} 
\end{aligned}
\\
\label{eq:transl_laci}
& \frac{d\, LacI}{dt}= \kappa_\mathrm{L}^\mathrm{p} \cdot mRNA_{\mathrm{LacI}} - g_\mathrm{L}^\mathrm{p} \cdot LacI\\
\label{eq:transl_tetr}
& \frac{d\, TetR}{dt}= \kappa_\mathrm{T}^\mathrm{p} \cdot mRNA_{\mathrm{TetR}} - g_\mathrm{T}^\mathrm{p} \cdot TetR
\end{align}
\normalsize
In the above equations, the state variables represent the concentration of mRNAs and proteins of the \emph{LacI} and \emph{TetR} promoters. 
The parameters $\kappa_\mathrm{L/T}^\mathrm{m0}$, $\kappa_\mathrm{L/T}^\mathrm{m}$, $\kappa_\mathrm{L/T}^\mathrm{p}$, $g_\mathrm{L/T}^\mathrm{m}$, and $g_\mathrm{L/T}^\mathrm{p}$ are the  transcription, translation, mRNA degradation, and protein degradation rates.
The inducer molecules $aTc$ and $IPTG$ influence the mRNA transciption rates through Hill functions that depend on the regulation parameters $\theta_\mathrm{aTc}, \theta_\mathrm{IPTG}, \theta_\mathrm{LacI}, \theta_\mathrm{TetR}$ and $\eta_\mathrm{aTc}, \eta_\mathrm{IPTG}, \eta_\mathrm{LacI}, \eta_\mathrm{TetR}$. 
All parameter values are the same as those used in \cite{Lugagne2017} and are also provided in Table \ref{tab:Parameters}.

The inducer molecules diffuse across the cell membrane with non-symmetrical dynamics as described in \cite{Lugagne2017}. Specifically, we have

\footnotesize
\begin{align}
\label{eq:diffusion_atc}
\frac{d\, aTc}{dt}= &
\begin{cases}
k^{\mathrm{in}}_{\mathrm{aTc}} (u_{\mathrm{aTc}} - aTc), & \mbox{ if }\ u_{\mathrm{aTc}} > aTc\\
k^{\mathrm{out}}_{\mathrm{aTc}} (u_{\mathrm{aTc}} - aTc), & \mbox{ if }\ u_{\mathrm{aTc}} \leq aTc
\end{cases},\\
\label{eq:diffusion_iptg}
\frac{d\, IPTG}{dt}= &
\begin{cases}
k^{\mathrm{in}}_{\mathrm{IPTG}} (u_{\mathrm{IPTG}} - IPTG), & \mbox{ if }\ u_{\mathrm{IPTG}} > IPTG\\
k^{\mathrm{out}}_{\mathrm{IPTG}} (u_{\mathrm{IPTG}} - IPTG), & \mbox{ if }\ u_{\mathrm{IPTG}} \leq IPTG
\end{cases},
\end{align}
\normalsize 
where $aTc$ and $IPTG$ denote the concentrations of the inducer molecules inside the cell, while $u_{\mathrm{aTc}}$ and $u_{\mathrm{IPTG}}$ those in the growth medium and represent the inputs to our system.
Under certain hypotheses on the ratio between the concentration of the external inducers, the system shows a bistable dynamics \cite{Gardner2000}. Conversely, when the ratio between the external inducers reaches a certain value, a saddle-node bifurcation occurs and the system becomes monostable.

Since the time scales of the mRNA dynamics is notably faster than that of the proteins, as discussed in \cite{Fiore2018}, by setting $\frac{d\, mRNA_{\mathrm{LacI}}}{dt}=0$ and  $\frac{d\, mRNA_{\mathrm{TetR}}}{dt}=0$ we can obtain the nondimensional Quasi-Steady State Model (see \cite{Fiore2018} for the derivation):
\begin{equation}
\label{eq:qssmodel}
\begin{split}
\frac{dx_1}{dt'} &= k_1^0 + \frac{k_1}{1+ x_2^2 \cdot w_1(t'/g^\mathrm{p}) }  - x_1\\
\frac{dx_2}{dt'} &= k_2^0 + \frac{k_2}{1+ x_1^2 \cdot w_2(t'/g^\mathrm{p}) }  - x_2
\end{split}
\end{equation}
where 
$$
t'=g^\mathrm{p}\, t, \ \ x_1=\frac{LacI}{\theta_{\mathrm{LacI}} }, \ \ x_2=\frac{TetR}{\theta_{\mathrm{TetR}}},
$$
and the adimensional parameters are defined as
$$
k_1^0=\frac{\kappa_\mathrm{L}^\mathrm{m0}\,\kappa_\mathrm{L}^\mathrm{p} }{g_\mathrm{L}^\mathrm{m}\,\theta_{\mathrm{LacI}}\, g^\mathrm{p}  }, \quad k_1=\frac{ \kappa_\mathrm{L}^\mathrm{m}\,\kappa_\mathrm{L}^\mathrm{p}}{g_\mathrm{L}^\mathrm{m}\,\theta_{\mathrm{LacI}}\, g^\mathrm{p}  },
$$ 
$$
k_2^0=\frac{\kappa_\mathrm{T}^\mathrm{m0}\,\kappa_\mathrm{T}^\mathrm{p} }{g_\mathrm{T}^\mathrm{m}\,\theta_{\mathrm{TetR}}\, g^\mathrm{p}  }, \quad k_2=\frac{\kappa_\mathrm{T}^\mathrm{m}\,\kappa_\mathrm{T}^\mathrm{p} }{g_\mathrm{T}^\mathrm{m}\,\theta_{\mathrm{TetR}}\, g^\mathrm{p}  }.
$$

The inputs to the system are modelled by the nonlinear functions $w_1$ and $w_2$ defined as:
\begin{align*}
w_1(aTc(t))= &  \frac{1}{\left(1 + \left( \frac{aTc(t)}{\theta_{\mathrm{aTc}}} \right)^{\eta_{\mathrm{aTc}}} \right)^{\eta_{\mathrm{TetR}}} } \\
w_2(IPTG(t))= & \frac{1}{\left(1 + \left( \frac{IPTG(t)}{\theta_{\mathrm{IPTG}}} \right)^{\eta_{\mathrm{IPTG}}} \right)^{\eta_{\mathrm{LacI}}}}.
\end{align*}
A schematic of the Toggle Switch model \eqref{eq:qssmodel} represented as a MIMO system is presented in \figurename~\ref{fig:GTSMIMO}. 
Therein, the input variables, namely $aTc$ and $IPTG$, represent the concentrations of the inducers inside the cell and the output variables $y_1$ and $y_2$ are fluorescent markers proportional to the concentration of the proteins, RFP for LacI and GFP for TetR, given by
\begin{equation}\label{eq:output}
	\begin{cases}
	y_1=k_\mathrm{RFP} \cdot LacI\\
	y_2=k_\mathrm{GFP} \cdot TetR.
	\end{cases}
\end{equation}

\begin{table}[ht]
\renewcommand{\arraystretch}{1.3}
\centering
\begin{tabular}{|c|c||c|c||c|c|}
\hline
$k^\mathrm{m0}_\mathrm{L}$ & 3.20e-2 & $g^\mathrm{m}_\mathrm{L}$ & 1.386e-1 & $\theta_\mathrm{LacI}$ & 31.94 \\ \hline
$k^\mathrm{m0}_\mathrm{T}$ & 1.19e-1 & $g^\mathrm{m}_\mathrm{T}$ & 1.386e-1 & $\theta_\mathrm{IPTG}$ & 9.06e-2\\ \hline
$k^\mathrm{m}_\mathrm{L}$ & 8.30 & $g^\mathrm{p}_\mathrm{L}$ & 1.65e-2 & $\theta_\mathrm{TetR}$ & 30.00\\ \hline
$k^\mathrm{m}_\mathrm{T}$ & 2.06 & $g^\mathrm{p}_\mathrm{T}$ & 1.65e-2 & $\theta_\mathrm{aTc}$ & 11.65\\ \hline
$k^\mathrm{p}_\mathrm{L}$ & 9.726e-1 & $k^\mathrm{in}_\mathrm{IPTG}$ & 2.75e-2 & $\eta_\mathrm{LacI}$ & 2.00\\ \hline
$k^\mathrm{p}_\mathrm{T}$ & 1.170& $k^\mathrm{out}_\mathrm{IPTG}$ & 1.11e-1 & $\eta_\mathrm{IPTG}$ & 2.00\\ \hline
&&$k^\mathrm{in}_\mathrm{aTc}$ & 1.62e-1 & $\eta_\mathrm{TetR}$ & 2.00 \\ \hline
&&$k^\mathrm{out}_\mathrm{aTc}$ & 2.00e-2 & $\eta_\mathrm{aTc}$ & 2.00\\ \hline
\end{tabular}
\caption{Value of the parameters of the Model}
\label{tab:Parameters}
\end{table}
\section{Control Design}
\label{sec:control}
In this section we present three control strategies to regulate the expression level of the toggle switch to an arbitrary intermediate value.
As input signals we use pulse waves like those recently proposed in \cite{Lugagne2017}. Therein the authors showed both in-silico and in-vivo that such class of input signals has beneficial effects on the level of coherence of the cell population response. This effect has been also recently analyzed in \cite{Khammash2018}.

All stochastic simulations presented in this section were obtained using the Gillespie's Stochastic Simulation Algorithm (SSA) \cite{Gillespie1977} to accurately take into account the intrinsic biochemical noise of the cells \cite{Elowitz2002}.

\subsection{PI control at the population level}
\begin{figure}[!t]
    \centering
    \subfigure[PI at the Population Level with Continuous Inputs.]{
    \psfragfig[width=0.95\linewidth]{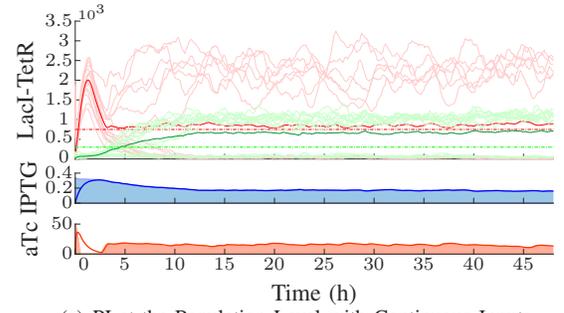}\llap{
    \parbox[b]{7.1cm}{\tiny $10^3$\\\rule{0ex}{3.65cm}}}
    \label{fig:PI_npa}}\\
    \subfigure[Pulse Wave parameters: $\bar{u}_\mathrm{IPTG}=0.5$, $\bar{u}_\mathrm{aTc}=50$, $T=100$ min.]{
    \psfragfig[width=0.95\linewidth]{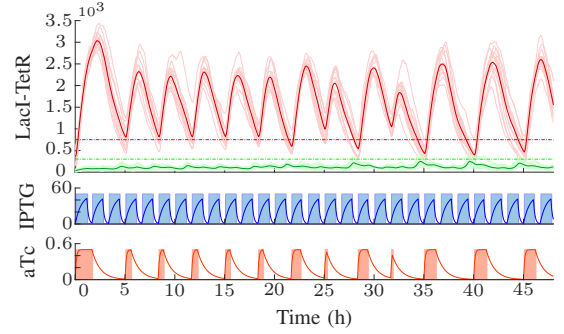}\llap{
    \parbox[b]{7.1cm}{\tiny $10^3$\\\rule{0ex}{4.03cm}}}
    \label{fig:PIPA}}
\caption{Performance of the PI controllers closing the loop on the average of the reporters' fluorescence over the population. The PIs' gains were tuned heuristically to $k^1_\mathrm{P}=0.05$, $k^1_\mathrm{I}=4\cdot 10^{-4}$, $k^2_\mathrm{P}=0.025$ and $k^2_\mathrm{I}=6.94\cdot 10^{-4}$. Simulation time is 48 hours and the population consists of 16 cells. Top panel in each subfigure: Evolution over time of $LacI$ (red) and $TetR$ (green): The trajectories of the single cells are reported as thin lines while their mean value over the population is reported as a thicker line. Central panel and bottom panel in each subfigure: Evolution over time of the concentrations of the inducer molecules inside (thick lines) and outside the cells (shaded areas).}
\label{fig:PI_controller}
\end{figure}

As a benchmark, we start by considering a control strategy consisting of two independent PI controllers, one per channel, whose respective loops are closed on the averages of the reporters' fluorescence over the population rather than on a single cell, as done in \cite{Lugagne2017}.
As reported in \figurename~\ref{fig:PI_npa}, we observe that, although the mean values of fluorescence are regulated in a neighbourhood of the target point, the population splits in two groups with cells converging onto either of the stable equilibria.

\begin{figure}[!t]
    \centering
    \includegraphics[width=0.9\linewidth]{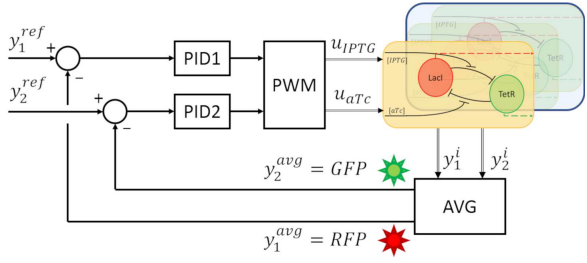}
    \caption{Block diagram of the PI control scheme closing the loop on the average of the reporters' fluorescence and modulating the control input via PWM.}
    \label{fig:MeanSchema}
\end{figure}

To reduce this high variance across the population, as suggested in \cite{Lugagne2017}, we exploit the benefits of using pulse wave inputs by modulating the control inputs generated by the PIs via PWM (\figurename~\ref{fig:MeanSchema}). This lead to higher coherence among cells (see \figurename~\ref{fig:PIPA}) but to higher regulation errors, too.


\subsection{Control via mutually exclusive pulse wave inputs}
\begin{figure}[tb]
	\begin{center}
		\includegraphics[width=\columnwidth]{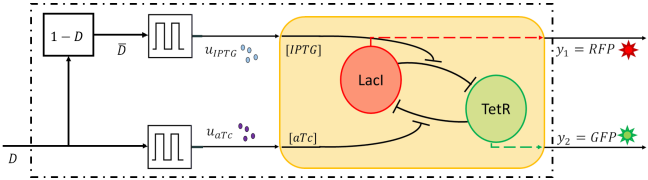}
		\caption{Single Input Multiple Output view of the system.}
		\label{fig:SIMO}
	\end{center}
\end{figure}

To overcome the problems of the PI approach presented above, we next modulate in a closed-loop manner two mutually exclusive pulse wave inputs as done in an open-loop experiment reported in \cite{Lugagne2017}.
Specifically, we choose
\begin{equation}\label{eq:sqwvinputs}
\begin{array}{ll}
u_\mathrm{aTc}(t)=\bar{u}_\mathrm{aTc}\cdot s_\mathrm{q}(t/T)\\
u_\mathrm{IPTG}(t)=\bar{u}_\mathrm{IPTG}\cdot(1-s_\mathrm{q}(t/T))
\end{array}
\end{equation}
with $s_\mathrm{q}(t/T)$ being a unitary pulse wave signal of period $T$ and duty-cycle $D\in[0,1]$, and $\bar{u}_\mathrm{aTc}$ and $\bar{u}_\mathrm{IPTG}$ their amplitudes. The system can then be seen as a SIMO - single input, multiple output - system, as shown in \figurename~\ref{fig:SIMO}.

During the synthesis, we assume instantaneous diffusion of the inducers across the cell membrane by setting: $$aTc(t)=u_\mathrm{aTc}(t)~\textrm{and}~IPTG(t)=u_\mathrm{IPTG}(t), \forall t\geq t_0.$$
We will then evaluate the effect of diffusion of the inputs across the cell membrane during validation in order to assess robustness of the control strategy.

With these assumptions on the inputs, we previously showed in \cite{Fiore2018} that, by applying nonlinear averaging techniques,  an \emph{average model} can be obtained that under some conditions captures the mean level of expression of the proteins over each period.
Specifically, by rescaling time setting $\tau=\frac{t'}{T g^\mathrm{p}}$ and integrating model \eqref{eq:qssmodel} over a period $T$, the toggle switch dynamics subject to \eqref{eq:sqwvinputs} can be described by the system (see \cite{Fiore2018} for details) 
\begin{equation}\label{eq:avgVF}
\begin{split}
\frac{d x_1}{d\tau} & =  \varepsilon \left[k_1^0+ k_1 \left( \frac{D}{1+x_2^2\cdot \bar{w}_1} + \frac{1-D}{1+x_2^2} \right) -x_1 \right]\\
\frac{d x_2}{d\tau} & =  \varepsilon \left[ k_2^0+ k_2 \left( \frac{D}{1+x_1^2} + \frac{1-D}{1+x_1^2 \cdot \bar{w}_2} \right) -x_2 \right]
\end{split}
\end{equation}
where $\varepsilon=Tg^\mathrm{p}$,
$\bar{w}_1=w_1(\bar{u}_\mathrm{aTc})$, and $\bar{w}_2=w_2(\bar{u}_\mathrm{IPTG})$.\\

\begin{figure}[tb]
    \centering
	\includegraphics[width=0.95\columnwidth]{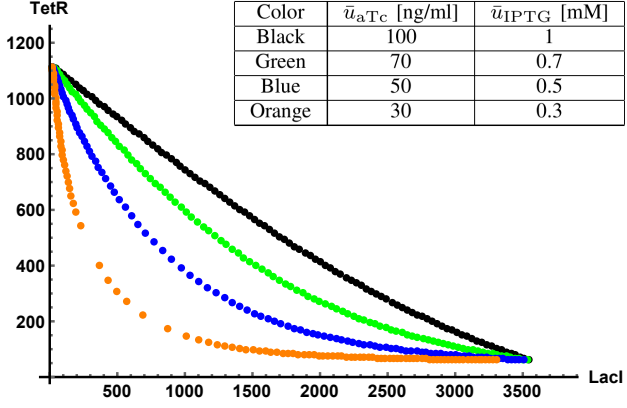}
	\caption{Curves of equilibrium points of system \eqref{eq:avgVF} as a function of the duty-cycle $D$ for different values of the input amplitudes as reported in the accompanying table. Each dot represents the position of the unique stable equilibrium of the system (\ref{eq:avgVF}) evaluated for $D$ in the interval $[0, 1]$ with increments of 0.01 for a given set of amplitude values. By varying the amplitude of the inputs, we obtain different curves in the plane; here, the ratio between the amplitude of the inputs is kept constant.}
	\label{fig:EqCurves}
\end{figure}

Note that, by varying the amplitudes $\bar{u}_\mathrm{aTc}$ and $\bar{u}_\mathrm{IPTG}$ and the duty-cycle $D$ of the inputs \eqref{eq:sqwvinputs}, the unique stable equilibrium point of system (\ref{eq:avgVF}) changes its location in state space so that different equilibrium curves, say $\Gamma_i({\bar{u}_\mathrm{aTc},\bar{u}_\mathrm{IPTG}})$, parameterized in $D$ can be obtained by varying the amplitudes of the input signals (see \figurename~\ref{fig:EqCurves} for some examples of such curves).

We will exploit model (\ref{eq:avgVF}) to synthesize two alternative feedback control strategies to select the amplitude of the input signals and vary on-line their duty-cycle to stabilize the toggle switch in a region where the genes are expressed at desired intermediate levels.

\subsubsection{\bf\em PI-PWM duty cycle compensation}
\begin{figure}[!t]
\centering
\includegraphics[width=\columnwidth]{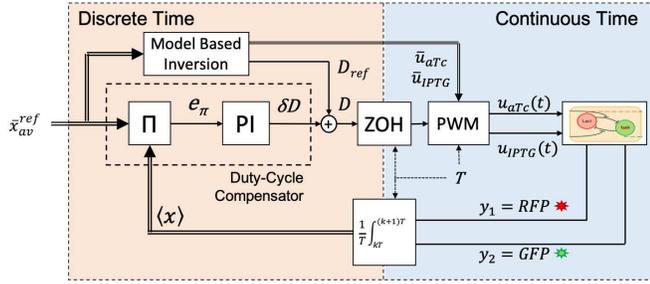}
\caption{Block diagram of the hybrid model-based PI-PWM approach.}
\label{fig:PIPWMSchema}
\end{figure}

The first approach we propose is a hybrid model-based strategy, whose block diagram is reported in \figurename~\ref{fig:PIPWMSchema}.
Here, the average model reported above is used to compute in feedforward the required amplitudes of the input signals, $\bar{u}_\mathrm{aTc}$ and $\bar{u}_\mathrm{IPTG}$, and the nominal value of the duty-cycle $D_{ref}$ required to achieve convergence of the toggle switch in a neighborhood of the desired values of the gene expression encoded by the reference signal. 

In absence of disturbances, diffusion dynamics and other unwanted effects, this feedforward action would itself suffice to achieve the control goal. To guarantee robustness when this unavoidable effects are present,  a discrete-time feedback PI action is also added to dynamically adjust at the end of each period the duty-cycle to compensate the mismatch between the average cell response predicted by the model and the one being measured.
We next describe in greater detail the key components of the control strategy shown in Fig. \ref{fig:PIPWMSchema}.
{\bf Model Based Inversion block.} This block uses model (\ref{eq:avgVF}) to select the amplitudes, $\bar{u}_\mathrm{aTc}$ and $\bar{u}_\mathrm{IPTG}$, and the nominal value of the duty cycle $D_{ref}$ that yield the equilibrium point closest to the desired reference target point $\bar{x}^{ref}_{av}$. 
To do so, a database of equilibrium curves such as those depicted in \figurename~\ref{fig:EqCurves} is interrogated. In our implementation, the database contains a total of 60 equilibrium curves, say $\Gamma_i({\bar{u}_\mathrm{aTc},\bar{u}_\mathrm{IPTG}})$, $i=1, \ldots, 60$, parameterized in $D_{ref}$ that were constructed by using the average model. 
The curves were obtained by considering values of
$\bar{u}_\mathrm{aTc} \in [0,100]$ with a step size of $5$ and $\bar{u}_\mathrm{IPTG} \in [0,1]$ with a step size of $0.05$. Specifically, 20 curves were obtained by varying $\bar{u}_\mathrm{IPTG}$ while keeping constant $\bar{u}_\mathrm{aTc}=100$, 20 curves by varying $\bar{u}_\mathrm{aTc}$ with  $\bar{u}_\mathrm{IPTG}=1$ and 20 by varying simultaneously $\bar{u}_\mathrm{aTc}$ and $\bar{u}_\mathrm{IPTG}$ while keeping their ratio constant. 
{\bf Projection block $\mathbf{\Pi}$.} Using the values of $\bar{u}_\mathrm{aTc}$, $\bar{u}_\mathrm{IPTG}$ and $D_{ref}$ computed in feedforward, the toggle switch evolution under PWM control inputs should converge towards a periodic orbit of period $T$ with average value equal to the equilibrium point selected by the model inversion block. In practice, disturbances, noise and diffusion effects will make this average value different from the predicted one. 

\begin{figure}[!t]
    \centering
    \includegraphics[width=0.7\columnwidth]{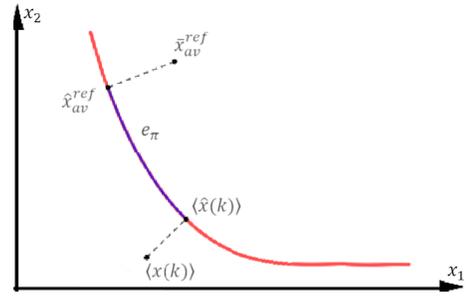}
    \caption{Evaluation of the projected error $e_{\pi}$. $\hat{x}_{av}^{ref}$ and $\langle \hat{x}(k)\rangle$ are, respectively the projections of $\bar{x}_{av}^{ref}$ and $\langle x(k) \rangle$ on the curve. The error $e_{\pi}$ is evaluated as the length of the arc between those two points. }
    \label{fig:projector}
\end{figure}

At the end of each period, the projection block $\Pi$ computes the error $e_{k,\pi}$ as the length of the arc on the selected equilibrium curve $\Gamma_{\bar{u}_\mathrm{aTc},\bar{u}_\mathrm{IPTG}}$ between the projections onto the same curve of the reference point $\bar{x}^{ref}_{av}$ and the current average state $\langle x(k)\rangle$  (\figurename~\ref{fig:projector}), evaluated as:
$$
\langle x(k)\rangle = \frac{1}{T} \int_{kT}^{(k+1)T} x(\tau)\, d\tau.
$$
 
{\bf PI Controller.} Finally, the error $e_{k,\pi}$ is compensated by a PI controller that evaluates the correction $\delta D_k$ to the duty-cycle at each period as:
\begin{equation*}
	\delta D_{k} = k_\mathrm{P}\,e_{k,\pi} + k_\mathrm{I} \sum_{j=0}^{k}e_{j,\pi},
\end{equation*}
so that the duty-cycle $D_{k}$ is then set as $D_{k} = D_{ref}+\delta D_{k}$, starting from $D_0=D_{ref}$. A zero-order hold (ZOH) is then used to close the loop via a pulse-width modulator (PWM).

\begin{figure*}[!t]
	\begin{center}
		\subfigure[Deterministic simulation of the toggle switch response in the absence of duty cycle compensation.]{\psfragfig[width=0.95\columnwidth]{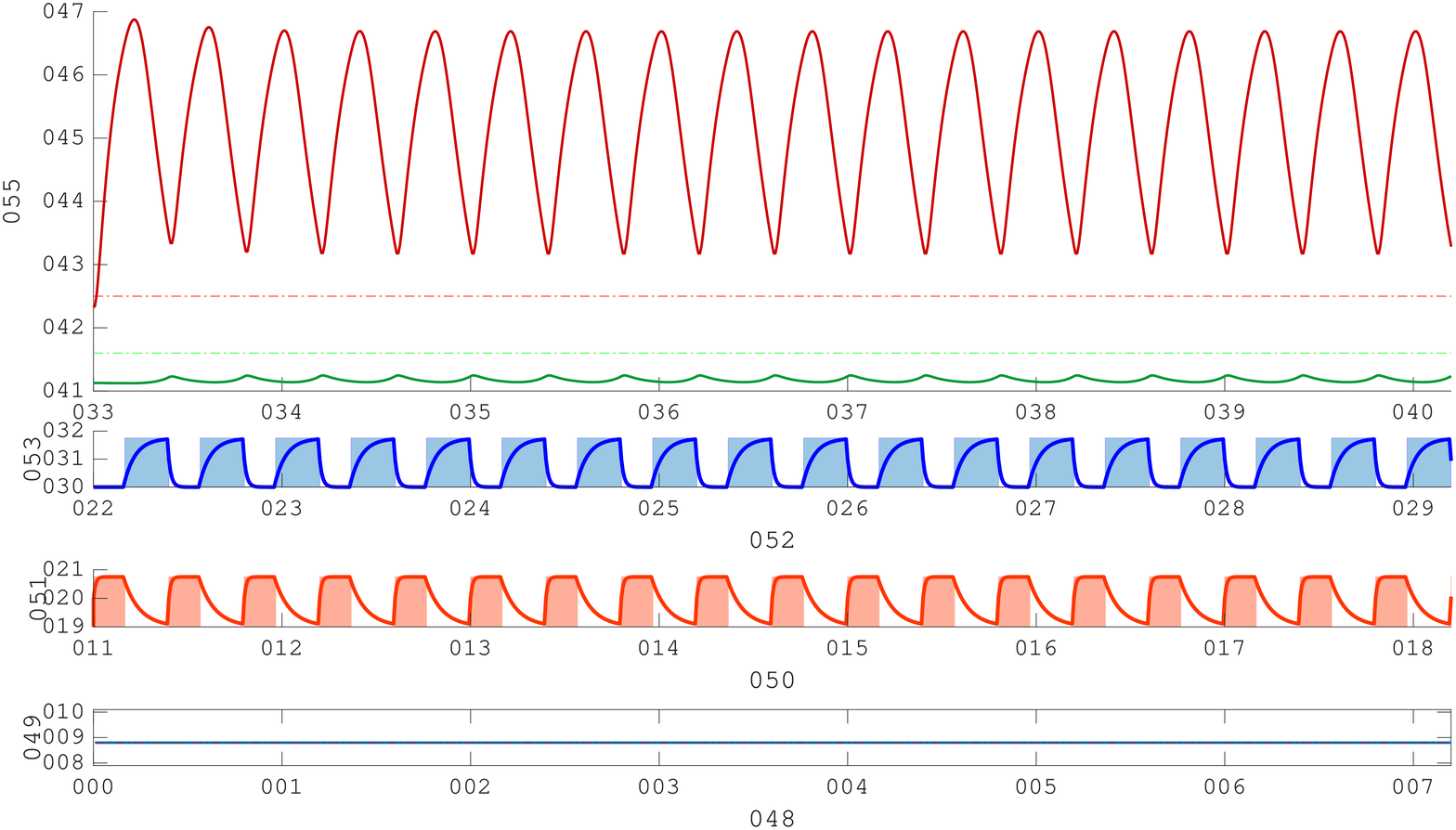}\llap{
        \parbox[b]{7.1cm}{\tiny $10^3$\\\rule{0ex}{4.3cm}}}
		\label{fig:PIPWModeOL}}
        \subfigure[Deterministic simulation of the system response under the action of the hybrid model-based PI-PWM feedback controller.]{\psfragfig[width=0.95\columnwidth]{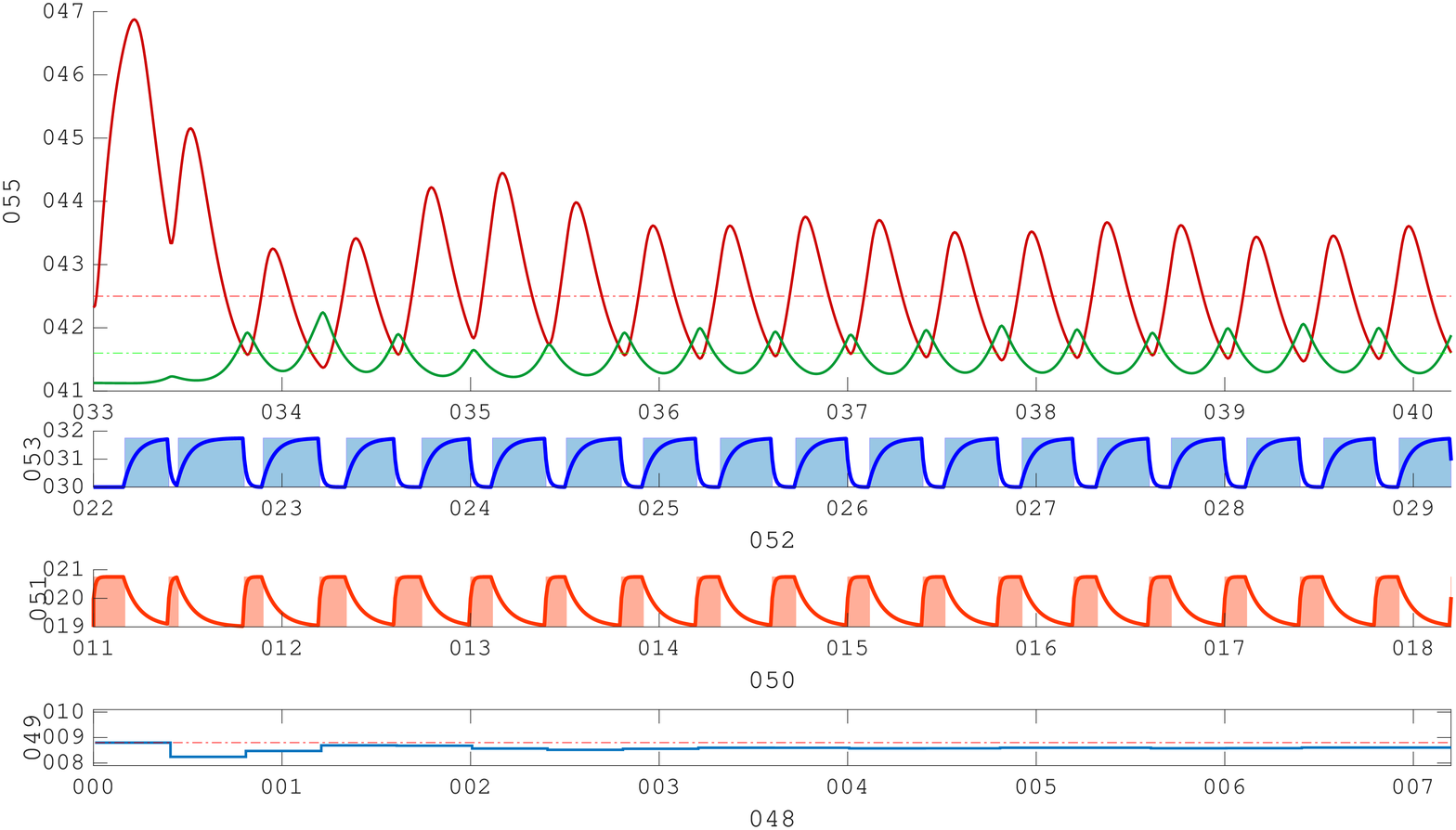}\llap{
        \parbox[b]{7.1cm}{\tiny $10^3$\\\rule{0ex}{4.3cm}}}
        \label{fig:PIPWModeCL}}
        \subfigure[Stochastic simulation of a 17 cell population. System response with complete PI-PWM feedback controller. The trajectory of the target cells are reported as thicker lines, while those of the others 16 cells are reported as thin lines.]{\psfragfig[width=0.95\columnwidth]{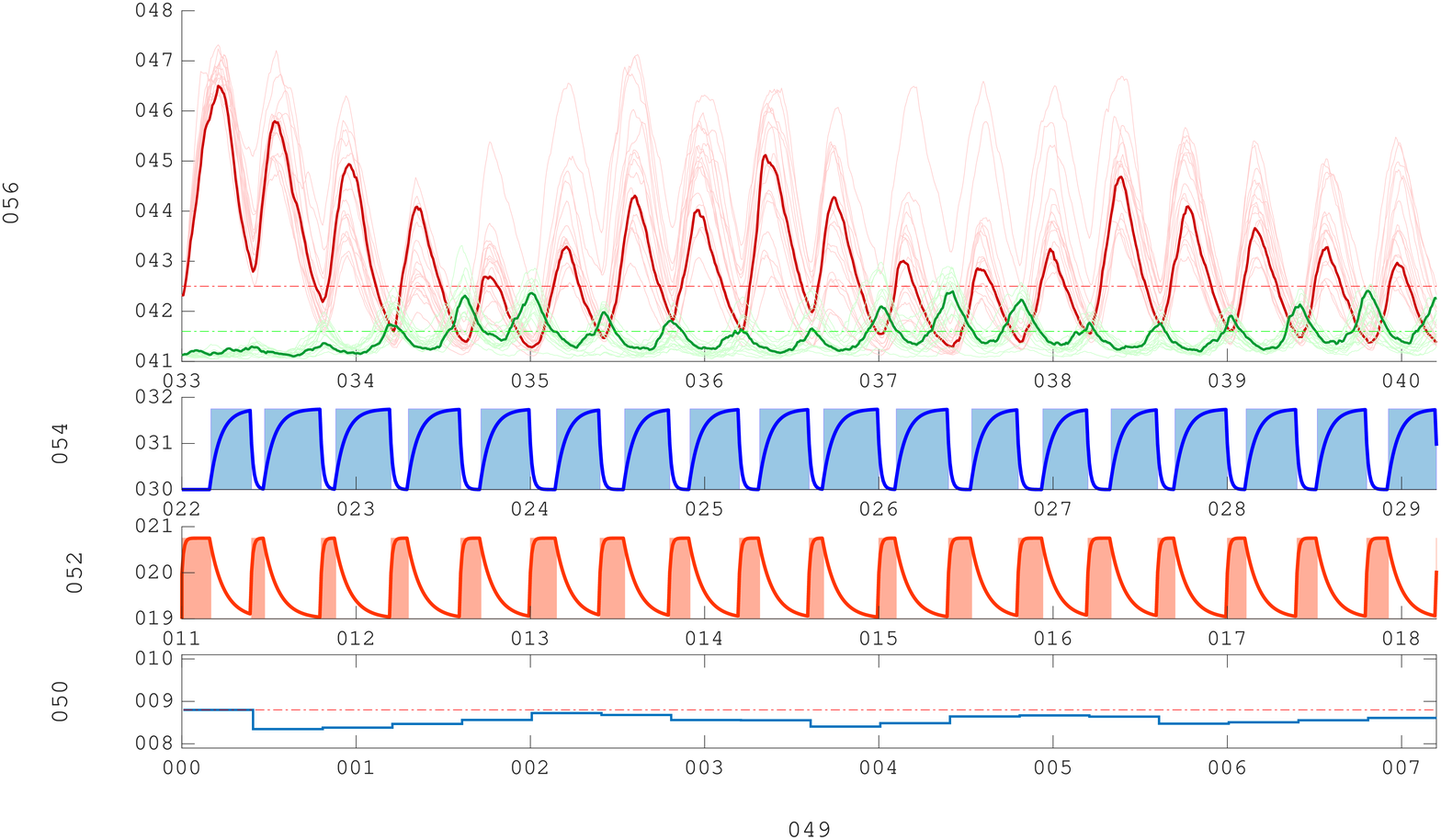}\llap{
        \parbox[b]{7.1cm}{\tiny $10^3$\\\rule{0ex}{4.3cm}}}
        \label{fig:PIPWMssa}}
        \subfigure[Mean trajectories over 10 stochastic simulation trials: The mean trajectories of the entire population over 10 simulations are reported as thick lines. The shaded areas represent the bounds containing all the trajectories obtained over 10 simulation trials.]{\psfragfig[width=0.95\columnwidth]{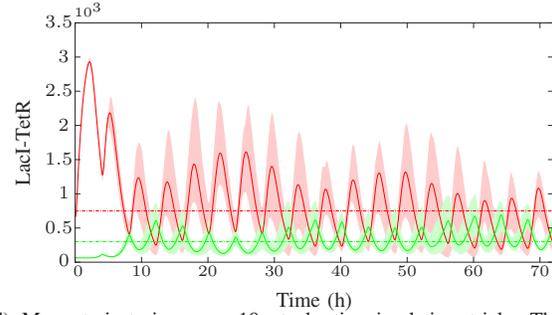}\llap{
        \parbox[b]{7.15cm}{\tiny $10^3$\\\rule{0ex}{3.7cm}}}
        \label{fig:PIPWMssa10m}}
	    \caption{PI-PWM technique. Simulation time is 72 hours. Pulse waves' parameters: $\bar{u}_\mathrm{IPTG}=0.35$, $\bar{u}_\mathrm{aTc}=35$, $T=240$ minutes. The PI's gains have been numerically tuned to reduce the settling time and the steady state error; their values are $k_\mathrm{p}=0.051$ and $k_\mathrm{i}=2.37\cdot 10^{-4}$. Reference setpoint is $LacI=750$, $TetR=300$; in relation to the system (\ref{eq:avgVF}), the setpoint is $\bar{x}_{av}^{ref}=[23.4821, 10.0002]$ that is projected onto the curve of equilibria at the point $\hat{x}_{av}^{ref}=[23.1100, 8.7173]$. \ref{fig:PIPWModeOL}-\ref{fig:PIPWMssa}: Top panel: Evolution of $LacI$ (red) and $TetR$ (green) over time. Central panels: Evolution of the concentrations of the inducer molecules inside and outside the cells, reported as thick line and shaded areas, respectively; Bottom panels: Duty-Cycle $D_k$ over time is reported in cyan, while the value $D_{ref}$ obtained from the Model based inversion block is reported as a red dashed line.}
	\end{center}
\end{figure*}

The performance of this controller was validated via both deterministic and stochastic simulations. As shown in \figurename~\ref{fig:PIPWModeOL}, when diffusion is present the open loop controller based on the use of solely the model inversion block is unable to guarantee convergence towards the desired values. Closing the loop instead makes the controller able to drive the toggle switch so that the expression levels of the proteins oscillate around the desired values with the duty-cycle being dynamically adjusted as expected.

Figure \ref{fig:PIPWMssa} shows a stochastic simulation for a population of 17 cells. Here the loop is closed onto a single cell (depicted by thicker red and green lines in the figure). We notice that the strategy is able to drive that cell towards the desired steady-state values but also to keep the standard deviation across the population contained in sharp contrast with the average PI controller whose performance has been shown in \figurename~\ref{fig:PI_controller}.
This is also confirmed by \figurename~\ref{fig:PIPWMssa10m} where the evolution of the cell population is averaged over 10 stochastic simulations.


\subsubsection{\bf\em ZAD}
As an alternative approach, we remove the need for model-based inversion by considering a different strategy, Zero Average Dynamics control -- a PWM-based control strategy originally presented in \cite{Fossas2000}. The technique has been developed to control electrical converters; however, its adaptability led to its application to several different fields, including Synthetic Biology \cite{Perrino2015}.

\begin{figure}[!t]
	\centering
	\includegraphics[width=\columnwidth]{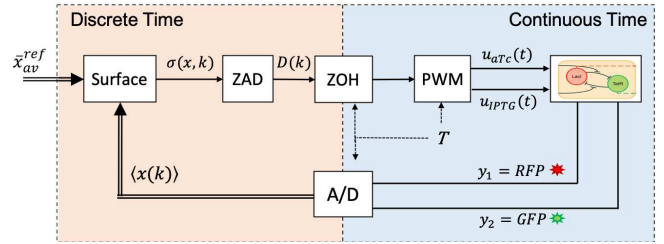}\\
	\caption{Block diagram of The Zero Average Dynamics controller.}
	\label{fig:ZADSchema}
\end{figure}
The control scheme is reported in \figurename~\ref{fig:ZADSchema}. 
The goal of the ZAD controller is to generate at each period $T$ two mutually exclusive pulse wave signals whose duty-cycle $D_k$ is such that the average value of some function $\sigma (x(t))$ is zero, that is
\begin{equation}
    \label{eq:zad_average}
    \mathbb{E}_{T}[\sigma (x(t))]=\int_{kT}^{(k+1)T} \!\! \sigma (x(\tau))\,d\tau = 0.
\end{equation}
By exploiting the fact that at regime there is a one-to-one correspondence of the average values of the state variables $LacI$ and $TetR$ (\figurename~\ref{fig:EqCurves}), we choose
$$
\sigma (x(t)) = x_\mathrm{1}(t) - x_\mathrm{1}^{ref}=\left(LacI(t)-LacI^{ref}\right) / {\theta_\mathrm{LacI}},
$$
where the dynamics of $x_1$ is described in \eqref{eq:qssmodel}, under the assumption of instantaneous diffusion.
Note that the monotonic dependence of the state variables at steady state with respect to the inputs $u_\mathrm{aTc}$ and $u_\mathrm{IPTG}$ also guarantees attractiveness of the surface defined by $\{\sigma(x)=0\}$.

By considering a piecewise-linear approximation of $\sigma(t)$ in each period, the duty-cycle is found as:
$$
D_{k}= 1- \sqrt{\dfrac{2\sigma_k+T\dot{\sigma}^\mathrm{on}_k}{T\left(\dot{\sigma}^\mathrm{on}_k-\dot{\sigma}^\mathrm{off}_k\right)}}
$$
where $\sigma_k=\sigma(kT)$ and
$$
\begin{aligned}
\dot{\sigma}^\mathrm{on}_k=\tfrac{d\sigma(x(kT))}{dt}\Bigg\vert_{u=[\bar{u}_\mathrm{aTc}, 0]}, \;\;
\dot{\sigma}^\mathrm{off}_k=\tfrac{d\sigma(x(kT))}{dt}\Bigg\vert_{u=[0, \bar{u}_\mathrm{IPTG}]}.
\end{aligned}
$$

\begin{figure}[!t]
	\centering
		\subfigure[Deterministic case with instantaneous diffusion across cell membrane.]{\psfragfig[width=0.95\columnwidth]{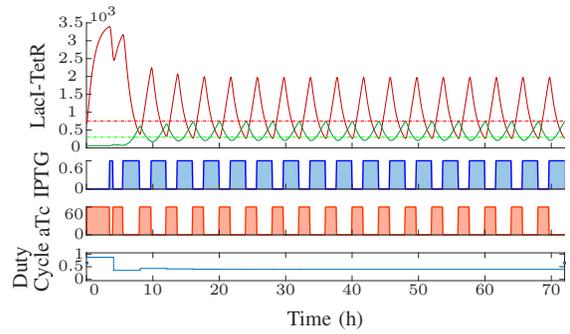}\llap{
        \parbox[b]{7.1cm}{\tiny $10^3$\\\rule{0ex}{4cm}}}
		\label{fig:ZADodeNoDiff}}\\
		\subfigure[Stochastic case with instantaneous diffusion across cell membrane, 17 cells. Top panel: the evolution of the target cell is reported as thick lines, while those of the other cells as thin lines.]{\psfragfig[width=0.95\columnwidth]{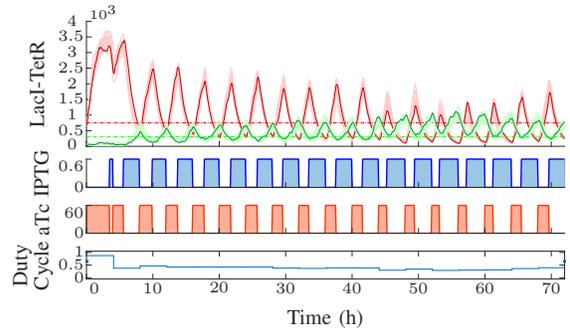}\llap{
        \parbox[b]{7.1cm}{\tiny $10^3$\\\rule{0ex}{4cm}}}
		\label{fig:ZADssaNoDiff}}\\
		\subfigure[Deterministic case with diffusion dynamics across cell membrane.]{\psfragfig[width=0.95\columnwidth]{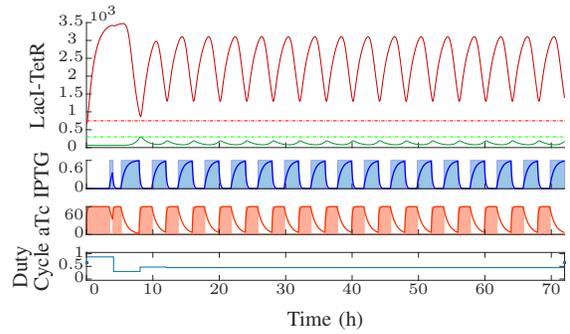}\llap{
        \parbox[b]{7.1cm}{\tiny $10^3$\\\rule{0ex}{4cm}}}
		\label{fig:ZADodeDiff}}
		\caption{ZAD control technique. Simulation time is 72 hours. Pulse waves parameters: $u_\mathrm{IPTG}=0.5$, $u_\mathrm{aTc}=50$, $T=240$ minutes. Top panel: Evolution of $LacI$ (red) and $TetR$ (green) over time. Central panels: 
		Evolution of the concentrations of the inducer molecules inside and outside the cells, reported as thick lines and shaded areas, respectively; Bottom panel: Duty-Cycle $D_k$ over time.}
\end{figure}

We validated the ZAD controller in both deterministic and stochastic settings by assuming instantaneous diffusion and then tested its robustness in the presence of diffusion of the control inputs through the cell membranes.
\figurename~\ref{fig:ZADodeNoDiff} shows that the strategy is effective when applied to control the deterministic system in the absence of diffusion with the stochastic simulation reported in \figurename~\ref{fig:ZADssaNoDiff} also confirming low standard deviation across the entire population.

However, when diffusion dynamics is added to the model, the ZAD controller is unable to guarantee a good performance as shown by the deterministic simulation reported in \figurename~\ref{fig:ZADodeDiff}. 
This is due to the fact that our implementation of the ZAD controller does not take diffusion explicitly into account in computing $D_k$. Indeed, the presence of diffusion introduces an undesirable delay in the system which disrupts the ability of the ZAD controller to achieve the control goal and would need to be appropriately compensated. This is the subject of ongoing work which will be reported elsewhere.

%
\section{Conclusions}
\label{sec:conclusion}
We discussed two feedback control strategies to stabilize a genetic toggle switch in a region where the expression levels of the two genes are intermediate. We started from the observation reported in \cite{Lugagne2017} that the use of mutually exclusive pulsing inputs can stabilize the switch in a region surrounding its unstable equilibrium guaranteeing limited standard deviation across the cell population being controlled. To overcome the unavoidable limitations of the open-loop control approach proposed therein, we exploited an average model of the toggle switch subject to this type of inputs to synthesize feedback control approaches able to adjust online the duty-cycle of the periodic inputs. 
We showed that a model-based PI-PWM strategy for duty-cycle compensation is effective in achieving the desired target region for the expression levels of the genes while also guaranteeing coherence of the fluorescence levels across the cell population. A second strategy based on the ZAD approach was shown to be viable when diffusion is not explicitly taken into account but exhibited limitations otherwise.
Future work will be aimed at improving our current ZAD implementation and developing an alternative approach based on MPC to select the duty-cycle of the periodic, mutually exclusive inputs. 

\footnotesize
\section*{ACKNOWLEDGMENT}

The authors wish to acknowledge support from the research project COSY-BIO (Control Engineering of Biological Systems for Reliable Synthetic Biology Applications) funded by the European Union's Horizon 2020 research and innovation programme under grant agreement No 766840.
%

\bibliographystyle{IEEEtran}
\bibliography{refs}

\begin{thebibliography}{10}
\providecommand{\url}[1]{#1}
\csname url@samestyle\endcsname
\providecommand{\newblock}{\relax}
\providecommand{\bibinfo}[2]{#2}
\providecommand{\BIBentrySTDinterwordspacing}{\spaceskip=0pt\relax}
\providecommand{\BIBentryALTinterwordstretchfactor}{4}
\providecommand{\BIBentryALTinterwordspacing}{\spaceskip=\fontdimen2\font plus
\BIBentryALTinterwordstretchfactor\fontdimen3\font minus
  \fontdimen4\font\relax}
\providecommand{\BIBforeignlanguage}[2]{{%
\expandafter\ifx\csname l@#1\endcsname\relax
\typeout{** WARNING: IEEEtran.bst: No hyphenation pattern has been}%
\typeout{** loaded for the language `#1'. Using the pattern for}%
\typeout{** the default language instead.}%
\else
\language=\csname l@#1\endcsname
\fi
#2}}
\providecommand{\BIBdecl}{\relax}
\BIBdecl

\bibitem{Gardner2000}
T.~S. Gardner, C.~R. Cantor, and J.~J. Collins, ``{Construction of a genetic
  toggle switch in Escherichia coli},'' \emph{Nature}, vol. 403, no. 6767, p.
  339, 2000.

\bibitem{Lugagne2017}
J.-B. Lugagne, S.~{Sosa Carrillo}, M.~Kirch, A.~K{\"{o}}hler, G.~Batt, and
  P.~Hersen, ``{Balancing a genetic toggle switch by real-time feedback control
  and periodic forcing},'' \emph{{Nature Communications}}, vol.~8, no.~1, p.
  1671, 2017.

\bibitem{Sootla2010}
A.~Sootla and D.~Ernst, ``{Pulse-based control using Koopman operator under
  parametric uncertainty},'' \emph{IEEE Transactions on Automatic Control},
  vol.~63, no.~3, pp. 791--796, 2018.

\bibitem{Sootla2016}
A.~Sootla, D.~Oyarz{\'u}n, D.~Angeli, and G.-B. Stan, ``{Shaping pulses to
  control bistable systems: Analysis, computation and counterexamples},''
  \emph{Automatica}, vol.~63, pp. 254--264, 2016.

\bibitem{Sootla2018}
A.~Sootla, A.~Mauroy, and D.~Ernst, ``{Optimal control formulation of
  pulse-based control using Koopman operator},'' \emph{Automatica}, vol.~91,
  pp. 217--224, 2018.

\bibitem{Sootla2013}
A.~Sootla, N.~Strelkowa, D.~Ernst, M.~Barahona, and G.-B. Stan, ``{Toggling a
  genetic switch using reinforcement learning},'' \emph{{Proc. of 9th french
  meeting on planning, decision making and learning}}, 2014.

\bibitem{Esfahani2013}
P.~M. Esfahani, ``{Analysis of Controlled Biological Switches via Stochastic
  Motion Planning},'' \emph{Proc. of the European Control Conference}, no.~1,
  pp. 93--98, 2013.

\bibitem{Chaves2011}
M.~Chaves and J.-L. Gouz{\'e}, ``Exact control of genetic networks in a
  qualitative framework: the bistable switch example,'' \emph{Automatica},
  vol.~47, no.~6, pp. 1105--1112, 2011.

\bibitem{Kapitza1951}
P.~Kapitza, ``Dynamic stability of a pendulum with an oscillating point of
  suspension,'' \emph{Journal of Experimental and Theoretical Physics},
  vol.~21, no.~5, pp. 588--597, 1951.

\bibitem{Fiore2018}
D.~Fiore, A.~Guarino, and M.~di~Bernardo, ``{Analysis and Control of Genetic
  Toggle Switches Subject to Periodic Multi-Input Stimulation},'' \emph{IEEE
  Control Systems Letters}, vol.~3, no.~2, pp. 278--283, 2019.

\bibitem{Khammash2018}
D.~Benzinger and M.~Khammash, ``Pulsatile inputs achieve tunable attenuation of
  gene expression variability and graded multi-gene regulation,'' \emph{{Nature
  Communications}}, vol.~9, no.~1, p. 3521, 2018.

\bibitem{Gillespie1977}
D.~T. Gillespie, ``Exact stochastic simulation of coupled chemical reactions,''
  \emph{The Journal of Physical Chemistry}, vol.~81, no.~25, pp. 2340--2361,
  1977.

\bibitem{Elowitz2002}
P.~S. Swain, M.~B. Elowitz, and E.~D. Siggia, ``Intrinsic and extrinsic
  contributions to stochasticity in gene expression,'' \emph{Proceedings of the
  National Academy of Sciences}, vol.~99, no.~20, pp. 12\,795--12\,800, 2002.

\bibitem{Fossas2000}
E.~Fossas, R.~Grin{\'o}, and D.~Biel, ``{Quasi-Sliding control based on pulse
  width modulation, zero averaged dynamics and the L2 norm},'' in
  \emph{{Advances In Variable Structure Systems: Analysis, Integration and
  Applications}}.\hskip 1em plus 0.5em minus 0.4em\relax World Scientific,
  2000, pp. 335--344.

\bibitem{Perrino2015}
G.~Fiore, G.~Perrino, M.~di~Bernardo, and D.~di~Bernardo, ``In vivo real-time
  control of gene expression: a comparative analysis of feedback control
  strategies in yeast,'' \emph{{ACS synthetic biology}}, vol.~5, no.~2, pp.
  154--162, 2015.

\end{thebibliography}

\end{document}